\documentclass[aps,prb,twocolumn,superscriptaddress,showpacs,showkeys]{revtex4}

\usepackage{graphicx,color,soul}
\newcommand{\urusi}{URu$_2$Si$_2$}

\newcommand{\etal}{{\itshape et al.}}

\newcommand{\pho}{\psi_{\rm HO}}
\newcommand{\paf}{\psi_{\rm AF}}
\newcommand{\qho}{\vec{Q}_{\rm HO}}
\newcommand{\qaf}{\vec{Q}_{\rm AF}}

\newcommand{\qzero}{(1,0,0)}
\newcommand{\qone}{(1.4,0,0)}

\newcommand{\pgnfigure}[2]{\begin{figure}\includegraphics[width=7cm]
{#1}\caption{\label{#1}#2}\end{figure}}
\newcommand{\pgnfigurestwo}[3]{\begin{figure}\includegraphics[width=7cm,clip=true]
{#1} \par \par\includegraphics[width=7cm,clip=true]{#2}
\caption{\label{#1}#3}\end{figure}}

\definecolor{mycolor}{rgb}{0.8, 0.2, 0.2}

\begin{document}

\title{Role of commensurate and incommensurate low-energy
excitations in the paramagnetic to hidden-order transition of
URu$_2$Si$_2$}

\author{P. G. Niklowitz}
\email[e-mail:]{philipp.niklowitz@rhul.ac.uk}
\affiliation{Department of Physics, Royal Holloway, University of
London, Egham TW20 0EX, UK}

\author{S. R. Dunsiger}
\affiliation{Center for Emergent Materials, Ohio State University, Columbus, OH 43210, USA}

\author{C. Pfleiderer}
\affiliation{Physik Department, Technische Universit\"at
M\"unchen, 85748 Garching, Germany}

\author{P. Link}
\affiliation{Heinz Maier-Leibnitz Zentrum (MLZ), Technische Universit\"at M\"unchen, Lichtenbergstr. 1, 85748
Garching, Germany}

\author{A. Schneidewind}
\affiliation{J\"ulich Centre for Neutron Science (JCNS) at Heinz Maier-Leibnitz Zentrum (MLZ), Forschungszentrum J\"ulich GmbH, Lichtenbergstr. 1, 85748 Garching, Germany}

\author{E. Faulhaber}
\affiliation{Heinz Maier-Leibnitz Zentrum (MLZ), Technische Universit\"at M\"unchen, Lichtenbergstr. 1, 85748
Garching, Germany}

\author{M. Vojta}
\affiliation{Institut f\"ur Theoretische Physik, Technische Universit\"at Dresden, 01062 Dresden, Germany}

\author{Y.-K. Huang}
\affiliation{Van der Waals-Zeeman Institute, University of Amsterdam, 1018XE Amsterdam, The Netherlands}

\author{J. A. Mydosh}
\affiliation{Kamerlingh Onnes Laboratory, Leiden University, 2300RA Leiden, The Netherlands}

\begin{abstract}
We report low-energy inelastic neutron scattering data of the
paramagnetic (PM) to hidden-order (HO) phase transition
at $T_0=17.5\,{\rm K}$ in URu$_2$Si$_2$.
While confirming previous results for the HO and PM phases,
our data reveal a pronounced wavevector dependence of
low-energy excitations {\em across} the phase transition. To analyze the energy scans
we employ a damped harmonic oscillator model containing a fit parameter $1/\Gamma$ which is expected to diverge at a second-order phase transition. Counter to expectations the excitations at $\vec{Q}_1\approx\qone$ show an abrupt step-like suppression of $1/\Gamma$ below $T_0$, whereas
excitations at $\vec{Q}_0=\qzero$, associated with large-moment antiferromagnetism (LMAF) under pressure,
show an enhancement and a pronounced peak of $1/\Gamma$ {\em at} $T_0$. Therefore, at the critical HO temperature $T_0$, LMAF fluctuations become nearly critical as well. This is the behavior expected of a ``super-vector'' order parameter with nearly degenerate components for the HO and LMAF leading to nearly isotropic
fluctuations in the combined order-parameter space.
\end{abstract}

\pacs{61.05.F-,62.50.-p,71.27.+a,75.30.Kz}
\keywords{hidden order, antiferromagnetism, inelastic neutron scattering}

\maketitle


\section{Introduction}

For nearly thirty years one of the most prominent unexplained properties of $f$-electron
materials has been the phase transition in {\urusi} at $T_0\approx17.5\,{\rm K}$ into a
state refered to as 'hidden order' (HO) \cite{pal85a,map86a,sch86a,myd11a} as the nature of the order parameter remains unknown. The discovery of the HO
was soon followed by the observation of a small antiferromagnetic moment (SMAF),
$m_s\approx 0.01-0.04$~$\mu_{\rm B}$ per U atom \cite{bro87a}, then believed to be an
intrinsic property of the HO. The observation of a large-moment antiferromagnetic phase
(LMAF) with $m_s\approx 0.4$~$\mu_{\rm B}$ \cite{ami99a}\ {\it under pressure} consequently
prompted intense theoretical efforts to connect the LMAF with the SMAF and the HO. However, $\mu$SR, NMR, Larmor and magnetic neutron diffraction experiments suggested that the apparent SMAF is a result of the presence of LMAF in a small sample volume fraction.\cite{mat03a,ama04a,nik10a,bou11a}. Studies of the pressure--temperature phase
diagram of {\urusi} consistently establish the existence of a bicritical point, which implies
that HO and LMAF break {\em different} symmetries
\cite{mot03a,uem05a,has08a,mot08a,nik10a}. To explain these properties, exotic
scenarios of the HO have been proposed, such as incommensurate orbital currents \cite{cha02a}, helicity
order \cite{var06a}, multipolar order \cite{kis05a,hau09a,tak12a,kha14a}, order due to dynamic symmetry breaking \cite{elg09a}, so-called hastatic order \cite{cha13a}, or spin-orbit density waves \cite{ris12a,das14a}. Indications of breaking of the 4-fold tetragonal in-plane symmetry at $T_0$ have motivated the consideration of a spin-nematic state \cite{fuj11a,oka11a,rig14a,ton14a}.

Inelastic neutron scattering has been essential for gaining microscopic insight into the nature
of the HO (see e.g. \cite{bro91a,bou10a}). The existence of commensurate and incommensurate
excitations in the HO at $\vec{Q}_0=\qzero$ and $\vec{Q}_1\approx\qone$, respectively, had been
known for quite a while \cite{bro87a}. The incommensurate excitations at $\vec{Q}_1$ are reported to exist in the HO and LMAF phase, with gaps of approximately 4\,meV and 8\,meV, respectively \cite{bou03b,vil08a}, and with a reduced gap \cite{bro91a,bou14a} or even gapless nature \cite{wie07a} in the PM phase. The closing of the gap has been quantitatively linked to the specific heat jump at $T_0$ \cite{wie07a}. In contrast, the commensurate excitations at $\vec{Q}_0$ have previously only been observed in the HO phase (gap approximately 2\,meV) and no critical behavior at $T_0$ has been reported \cite{vil08a,bou10a}. Therefore, the link between excitations at $\vec{Q}_0$ and $\vec{Q}_1$ and the HO has remained unclear.


In this paper we present compelling evidence that the {\em commensurate} fluctuations at
$\vec{Q}_0$ do not disappear right at $T_0$, but evolve in a way across the PM--HO transition, which allows us to interpret them as precursors of LMAF order. Further, the LMAF and HO fluctuations are clearly interrelated and $\qaf=(0,0,1)$ is the most likely HO wave vector, while the incommensurate fluctuations at $\vec{Q}_1$ are mere bystanders. This is the result of a direct and quantitative comparison of the excitations at $\vec{Q}_0$ and $\vec{Q}_1$, which have been studied in a single neutron scattering experiment. Detailed temperature scans across the HO--PM phase transition at low energies turn out to be an ideal way to visualize in the raw data the existence of a clear qualitative difference in the behavior at $\vec{Q}_0$ and $\vec{Q}_1$. At the incommensurate $\vec{Q}_1$ position the gap is filled abruptly at $T_0$ upon heating, i.e., the low-energy excitations are enhanced in a step-like fashion upon entering the PM phase. In contrast, scans at the commensurate $\vec{Q}_0$ position show that the low-energy excitations are enhanced across a considerable temperature range and peak at $T_0$.

A detailed analysis of energy scans confirms that
the fluctuations at $\vec{Q}_0$, which can be understood as precursors of LMAF order become {\em almost critical} at the PM--HO transition, in addition to the expected critical fluctuations of the hitherto unidentified HO parameter. This is not expected in a standard scenario of competing order parameters for HO and LMAF, which break different symmetries. However, it is consistent with nearly isotropic fluctuations of a super-vector order parameter as described in the discussion, which consists of components for both HO and LMAF. Isotropy in this order-parameter space would imply an emergent symmetry between both orders, which may be tested experimentally.



\section{Experimental}
The 2g single crystal studied was grown by means of an optical floating-zone technique at
the Amsterdam/Leiden Center. High sample quality was confirmed via X-ray diffraction and
detailed electron probe microanalysis. The mosaic spread is less than $1^{\circ}$. Samples prepared from this ingot showed good
resistance ratios (20 for the $c$ axis and $\approx$ 10 for the $a$ axis) and a high
superconducting $T_c\approx 1.5$\,K. The magnetization of the
large single crystal agreed very well with data shown in Ref.\,\onlinecite{pfl06a} and
confirmed the absence of ferromagnetic inclusions. Most importantly, in our neutron
scattering measurements we found an antiferromagnetic moment $m_s \approx 0.012~\mu_B$
per U atom \cite{nik10a}, which matches the smallest moment reported so far
\cite{ami07a}.


Inelastic neutron scattering measurements were carried out at the cold triple-axis
spectrometer PANDA at FRMII. The sample was mounted on a Cd-shielded Cu holder and
oriented with $(h0l)$ as the horizontal scattering plane. PANDA was used in W-configuration
with vertically and horizontally focusing monochromator and analyser and no
collimation. The final wavevector was kept fixed at 1.55~\AA$^{-1}$. Higher-order
harmonics were removed from the scattered beam by a liquid-nitrogen cooled Be filter and monitor correction for higher order neutrons was included. The
temperature evolution of the low-energy excitations of {\urusi} at commensurate
$\vec{Q}_0$ and incommensurate $\vec{Q}_1$ was studied by low-energy scans at
different temperatures. Most importantly, detailed temperature scans at $E=0.5$~meV
were carried out at each position.

\pgnfigurestwo{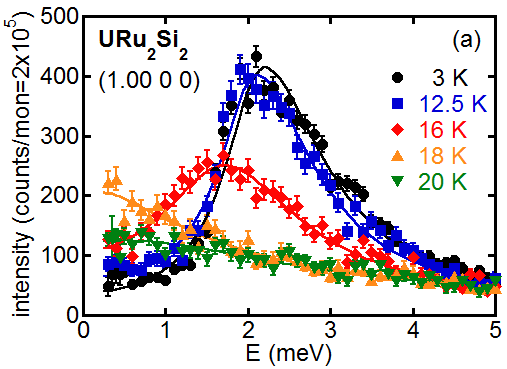}{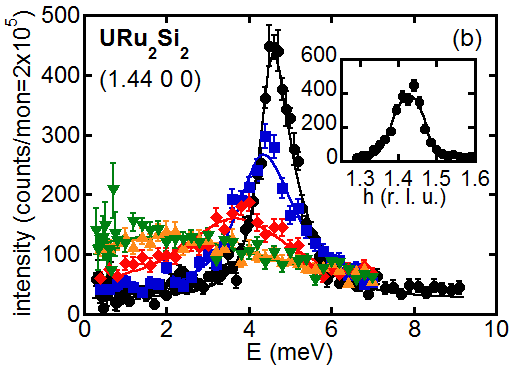}{
Low-energy excitations. At low temperature $T$ in the hidden order phase
excitations are gapped. Excitations are seen (a) above 2\,meV at the
commensurate $\qzero$ position and (b) above 4\,meV at the incommensurate
$(1.44,0,0)$ position. The inset shows a $h$ scan at 4.5\,meV at 3\,K. Both gaps close at the transition to paramagnetism ($T_0=17.5$\,K). (Counting time approximately 2\,min. Solid lines represent damped-harmonic oscillator fits described in the text.)
}

\subsection{Energy scans} 

Figure~\ref{figure1a.png}\ shows typical energy scans for
$\vec{Q}_0$ and $\vec{Q}_1$. At both positions scattering by considerably damped excitations is found for temperatures above $T_0$.
At $T_0$ gaps are opening up, and with further decreasing $T$ the intensities of the excitations increase
while the gaps widen. The spectrum is clearly gapped at low temperatures.
At 3~K, low-energy excitations are detected at 2~meV at $\vec{Q}_0$ and just above 4~meV
at $\vec{Q}_1$.

\pgnfigure{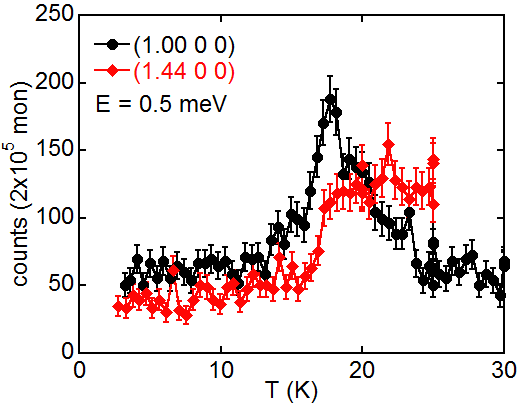}{
Temperature dependence of the low-energy excitations around the
hidden-order to paramagnetic phase transition. At the
incommensurate $(1.44,0,0)$ wavevector the gap is filled much more abruptly at $T_0$ and
the $E=0.5$\,meV excitations do not show any additional enhancement. At the commensurate $\qzero$ position the
gap is filled across a considerable temperature range around $T_0$. Low-energy
excitations at $E=0.5$\,meV are strongly enhanced and peak precisely at $T_0$ illustrating the peaking of the damping, which is exclusive to the $\qzero$ position.}

\subsection{Temperature scans}

While the spin fluctuations are truly critical neither at
$\vec{Q}_0$ nor at $\vec{Q}_1$, we have observed significant differences between the
low-energy excitation spectra at $\vec{Q}_0$ and $\vec{Q}_1$ when approaching the onset
of hidden order at $T_0$. For both wave vectors these differences are best visualized in detailed
temperature scans at $E=0.5$~meV (Fig.~\ref{figure2.png}).

At the incommensurate position, $\vec{Q}_1$, the gap opens in a step-like fashion at $T_0$, with the scattering intensity being essentially constant above $T_0$. In comparison to related data reported
by Wiebe \etal\ \cite{wie07a} for an energy transfer of $E=0.25$~meV, our scan at $E=0.5$~meV shows a much sharper decrease of scattering intensity just below $T_0$. This may be due to larger background contributions at the smaller energy transfer studied by Wiebe \etal. However, both data sets agree in that there is no enhancement of the
low-$E$ excitations when approaching $T_0$ from above.

In strong contrast, at the commensurate
position, $\vec{Q}_0$, the intensity of the low-energy excitations increases below about 24~K when approaching $T_0$, peaking precisely at $T_0$. It then decreases in a broad temperature range below $T_0$, becoming fully suppressed below $\sim 11$~K.

\pgnfigure{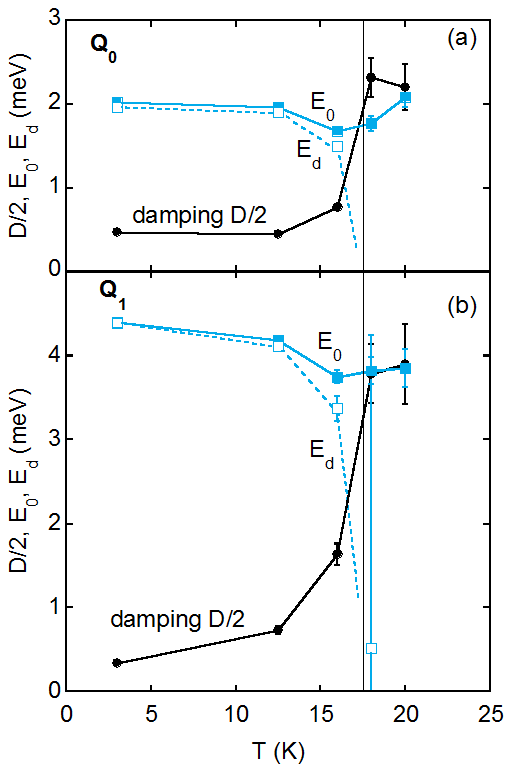}{Temperature dependence of the resonance energies $E_0$ of the undamped and $E_d$ of the damped oscillator and $D/2$ with damping $D$. Vertical lines indicate $T_0$ of the HO-PM transition. The behavior at (a) the commensurate position $\vec{Q}_0$ and (b) the incommensurate position $\vec{Q}_1$ is qualitatively similar. $E_0$ is fairly constant and shows a small dip near $T_0$, while $E_d$ is strongly suppressed near $T_0$. $D$ shows a significant increase near $T_0$. The excitations are weakly damped at low temperatures and approximately critically damped ($E_0\approx D/2$) above $T_0$. The increased errors at $T>T_0$ reflect that $E_0$ and $D$ and in particular $E_d$ become less well defined near the critically damped regime.}

\pgnfigure{figure4.png}{Temperature dependence of (a) the static susceptibility $\chi_0(\vec{q})=\chi(\vec{q},\omega=0)$ and of (b) $1/\Gamma$ with $\Gamma=E_0^2/D$. Vertical lines indicate $T_0$ of the HO--PM transition. (a) $\chi_0(\vec{Q}_0)$ drops while $\chi_0(\vec{Q}_1)$ rises with increasing temperature. $\chi_0(\vec{Q}_0)$ and $\chi_0(\vec{Q}_1)$ have similar magnitude near $T_0$. To allow this quantitative comparison of both signals $\chi_0$ values have been normalised by the form factor of UO$_2$ \cite{fra65a}. (b) $1/\Gamma(\vec{Q}_1)$ shows an almost step-like increase near $T_0$. $1/\Gamma(\vec{Q}_0)$ is larger than $1/\Gamma(\vec{Q}_1)$ in the whole studied temperature range and shows a strong peak at $T_0$. $\Gamma$ would correspond to the quasielastic linewidth in the strongly overdamped limit.}

\subsection{Analysis of energy scans}

The low-energy scattering in Figure~\ref{figure1a.png}\ consists of the contribution from
the dynamic structure factor $S(\vec{q},E)$ and a background. A main ingredient of $S(\vec{q},E)$ is the imaginary part $\chi''(\vec{q},E)$ of the dynamic susceptibility.
In previous studies of the commensurate excitations the imaginary part of the dynamic susceptibility was either modelled as a damped harmonic oscillator \cite{mas95b,bou10a} or, for $T>T_0$, in a  quasielastic approximation \cite{bou10a}. Moreover, in a different study, the incommensurate excitations were analyzed below and above $T_0$ with a Lorentzian model \cite{bro91a}. In comparison, we consistently fit, for the first time, our data at all temperatures at both $\vec{Q}_0$ and $\vec{Q}_1$ using a damped harmonic oscillator function for $\chi''(\vec{q})$:
\[S(\vec{q},E)=\left(\frac{1}{1-e^{-E/k_BT}}\right)\frac{\chi_0'(\vec{q})E_0^2D_{\vec{q}}E}{(E^2-E_0^2)
^2+D_{\vec{q}}^2E^2}\mbox{ .}\]
Here, $\chi_0'(\vec{q})$ is the real part of the static susceptibility, $D_{\vec{q}}$ denotes damping, and $E_0$ is the resonance energy of the undamped oscillator. $\left(\frac{1}{1-e^{-E/k_BT}}\right)=n_E+1$ is the detailed balance factor and contains the Bose-Einstein distribution $n_E=\frac{1}{e^{E/k_BT}-1}$. The detailed balance factor describes the temperature dependence of the probability for neutrons to scatter as a function of energy loss $E$. In particular, it can be seen that the scattering probability remains significant even if $E$ is much larger than $k_BT$.

The data in Figure~\ref{figure1a.png}\ was fitted using a convolution of $S(\vec{q},E)$ with the resolution ellipsoid, where we have assumed a quadratic dispersion $E_0=E_0(\vec{Q}_i)+a_{ih}(Q_{ih}-Q_h)^2+a_{ik}(Q_{ik}-Q_k)^2+a_{il}(Q_{il}-Q_l)^2$ near both $\vec{Q}_i$ with $i=0,1$. We have set $\vec{Q}_1=(1.41,0,0)$ as this is at the boundary of the body-centred tetragonal reciprocal lattice, on which the dispersion minimum seems to be centred \cite{but15a}. We infer dispersion parameters $a_{ih}$, $a_{ik}$ and $a_{il}$ from the results of Broholm \etal \cite{bro91a}. As the dispersion near $\vec{Q}_1$ in the $l$ direction was not reported in the literature, we have assumed $a_{1l}/a_{1h}=a_{0l}/a_{0h}$. We have assumed $\chi_0'$ to have a Gaussian $\vec{q}$ dependence in the immediate vicinity of both $\vec{Q}_i$ with $i=0,1$. We have obtained the strength of the dependence by taking the $\vec{q}$ dependence of the integrated excitation intensity obtained by Broholm et al. \cite{bro91a} at low temperatures as an order of magnitude estimate. For our fits, the resolution ellipsoid was determined in RESCAL \cite{resca} with the Cooper-Nathan method with values for the beam divergences derived from the instrument geometry. Fits were obtained by Monte Carlo integrations using Mfit4 \cite{resca}. 

We find that the background, as determined in energy scans at 20 and 3~K at (1.2 0 0), is
constant in the $E$ range of our experiment and weakly $T$-dependent. The
background is thereby assumed to include the magnetic continuum previously reported in Ref.\,\onlinecite{bou10a}. The fits shown in Figure~\ref{figure1a.png}\ are in excellent agreement with the data, thus supporting the suitability of the model.

The temperature evolution of $E_0$ and $D$ as well as the associated resonance energy of the damped oscillator, $E_d=\sqrt{E_0^2-(D/2)^2}$, are shown in Figure~\ref{figure3.png}. The condition for critical damping is $E_0=D/2$, which implies $E_d=0$. The behavior of $E_0$, $D$, and $E_d$ at $\vec{Q}_0$ and $\vec{Q}_1$ is qualitatively similar.  $E_0$ is fairly constant with values close to 2\,meV at $\vec{Q}_0$ and close to 4\,meV at $\vec{Q}_1$. In both cases a small dip is seen near $T_0$. $D$ strongly increases near $T_0$ and the damping level changes from moderate below $T_0$ to critically damped above $T_0$ in both cases. Near $T_0$, $E_d$ is suppressed to zero within the error in both cases.

When comparing our results at $\vec{Q}_0$ with the results of previous studies by Mason \etal \cite{mas95b} and Bourdarot \etal \cite{bou10a} the values for $E_0$ well below $T_0$ are consistent with each other and a qualitatively similar increase of $D$ near $T_0$ is found in each study. However, our observation of only a small dip of $E_0$ near $T_0$ is in contrast to the previously reported strong suppression of $E_0$ near $T_0$ and to the related equally strong suppression of the resonance energy in a magnetic excitation model \cite{bou14a}.

For $\vec{Q}_1$ comparison with other reports is more difficult, as either a three-parameter Lorentzian \cite{bro87a}\ or magnetic excitation model \cite{bou14a}\ was used. Our results for the temperature dependencies of $E_0$ and $D$ are qualitatively similar to the results found for the resonance energies and damping parameters, respectively, of the previously used models. However, the decrease of $E_0$ near $T_0$ in our analysis is comparatively small.

The most striking difference between our study and previous work concerns the $T$ dependence of the resonance energies at $\vec{Q}_0$ as compared to $\vec{Q}_1$ across $T_0$. While Ref.\,\onlinecite{bou14a} concludes on qualitative differences, we find qualitative similarity between $\vec{Q}_0$ and $\vec{Q}_1$. 

We attribute the discussed discrepancies between $E_0$ and $D$ values for $\vec{Q}_0$ and $\vec{Q}_1$ reported here and in the literature to the following effect: in the regime of stronger damping the harmonic oscillator function only depends on the combination $\Gamma=E_0^2/D$ except at higher $E$, where the function is close to zero. Therefore, $E_0$ and $D$ become less well defined as individual fit parameters in the regime of stronger damping. However, $\Gamma$ itself and the static susceptibility $\chi_0$ are well-defined parameters at any damping level and more appropriate for a comparison of the excitations at $\vec{Q}_0$ and $\vec{Q}_1$. We stress that $\Gamma$ can only be interpreted as the quasielastic linewidth or relaxation rate of the magnetic excitations in the strongly overdamped limit. Nevertheless, $\Gamma$ is a meaningful quantity at all damping regimes, as it is true for all levels of damping that an increase in $1/\Gamma$ represents a change to a more highly damped regime (as this depends on the ratio of $D$ and $E_0$).

Figure~\ref{figure4.png}a shows that with increasing $T$ the static susceptibility $\chi_0$ decreases at $\vec{Q}_0$ but increases at $\vec{Q}_1$. The $T$ dependence of $\chi_0(\vec{Q}_0)$ is similar to that reported in Ref.\,\onlinecite{bou10a}. A divergence as reported by Mason \etal \cite{mas95b} is not observed. A comparison of the behaviour of $\chi_0(\vec{Q}_1)$ with other studies is again more difficult due to the different models used in Refs.\,\onlinecite{bro87a} and \onlinecite{bou14a}. Nevertheless, the $T$ dependence of $\chi_0(\vec{Q}_1)$ is found to be qualitatively similar to the amplitude reported in Ref.\,\onlinecite{bro87a} and the static susceptibility reported in Ref.\,\onlinecite{bou14a}, which are the corresponding parameters of the respective models. To additionally allow for a quantitative comparison of the signals at $\vec{Q}_0$ and $\vec{Q}_1$ in this study, $\chi_0$ values have been normalised by the form factor of UO$_2$ \cite{fra65a}. $\chi_0(\vec{Q}_0)$ has a similar magnitude to $\chi_0(\vec{Q}_1)$ near $T_0$.

Figure~\ref{figure4.png}b shows that the $T$ dependence of $1/\Gamma$, like that of $\chi_0$, is also distinctly different at $\vec{Q}_0$ and $\vec{Q}_1$. In general, at a continuous phase transition,  $1/\Gamma$ of the order-parameter response function is expected to diverge. Experimentally, $1/\Gamma(\vec{Q}_0)$, although not diverging, displays a pronounced peak right at $T_0$. This indicates that the LMAF fluctuations become almost critical at the PM--HO transition, in addition to the expected critical fluctuations of the hitherto unidentified HO parameter. (We note that tiny amounts of quenched disorder may also limit $1/\Gamma(\vec{Q}_0)$ at $T_0$.)

However, the behavior of $1/\Gamma(\vec{Q}_1)$ is very different. With increasing $T$, $1/\Gamma(\vec{Q}_1)$ only shows a step-like evolution across $T_0$. Also, $1/\Gamma(\vec{Q}_1)<1/\Gamma(\vec{Q}_0)$ in the whole $T$ range studied. A comparison with the raw data in Figure~\ref{figure2.png} shows that $1/\Gamma$ captures the main difference between the $T$ dependencies of the low-energy excitation spectra at $\vec{Q}_0$ and $\vec{Q}_1$.




\section{Discussion}

Our experimental results suggest an intimate
relation between the PM--HO phase transition and the commensurate excitations at
$\vec{Q}_0$. The latter show enhanced damping towards $T_0$, much like the critical
fluctuations of a second-order magnetic phase transition. In marked contrast, the incommensurate
excitations at $\vec{Q}_1$ do not show a peak of $1/\Gamma$. Instead, $\Gamma$ remains
essentially constant when $T_0$ is approached from high temperatures. The
opening of the gap in these incommensurate excitations may in turn be interpreted
as a simple {\em consequence} of the onset of the HO. This view is not incompatible
with the proposal \cite{wie07a} that the incommensurate excitations at ${\vec Q}_1$ are mainly
responsible for the magnitude of the specific-heat anomaly.

The commensurate fluctuations at $\vec{Q}_0$
-- although being peaked exactly at $T_0$ -- do not become critical, i.e., the
corresponding static susceptibility does not diverge. The latter is
only expected if the magnetic order with $\qaf=(0,0,1)$, which is represented by a Bragg peak at $\vec{Q}_0$, becomes static below $T_0$. This would be the case in the pressure-induced LMAF phase, but not in the HO phase.
What is then the role of the $\vec{Q}_0$ magnetic fluctuations?

It is instructive to discuss the interplay of hidden order and magnetism using the
order-parameter language. If HO and magnetism would simply represent competing order
parameters $\pho$ and $\paf$, with ordering wavevectors $\qho$ and $\qaf$, respectively,
an enhancement of the HO would lead to a suppression of magnetism and vice versa. In
particular, it would be expected that magnetic fluctuations are suppressed instead of
enhanced when approaching the HO transition. Moreover, one would not expect that the HO
couples to the magnetism in a wavevector-selective manner: to lowest order the allowed
coupling in a Landau functional is of the form $|\pho|^2|\paf|^2$, which does not
require any relationship between $\qho$ and $\qaf$. Therefore, a standard scenario of
competing orders with differing symmetries, inferred from the parasitic nature of the
small-moment antiferromagnetism and the temperature-pressure phase diagram \cite{nik10a},
does not easily account for our data.

This prompts us to invoke a closer relationship between HO and LMAF. Specific
proposals along these lines were recently made in Refs.~\onlinecite{hau09a,hau09b} for hexadecapolar order and in Ref.~\onlinecite{cha13a} for hastatic order.
The central idea for such a closer relationship, common to these different microscopic calculations
\cite{hau09a,cha13a}, is that $\pho$ and $\paf$ may be treated as components of a {\em common}
super-vector order parameter. This implies that the system is in the vicinity
of a point with higher symmetry, where HO and magnetism are degenerate. Approaching the
ordering transition at $T_0$ in turn will lead to a concomitant enhancement of {\em both} HO and
magnetic fluctuations, corresponding to nearly isotropic fluctuations in
order-parameter space, until -- very close to the PM--HO transition -- the magnetic
fluctuations are cut-off, consistent with our data. In the simplest case, this scenario
suggests $\qho=\qaf=(0,0,1)$.
Therefore it will be crucial to search for these proposed order parameters \cite{hau09a,cha13a} at wavevector $(0,0,1)$ in the HO phase as well as their fluctuations for $T>T_0$.


\section{Conclusions}

We have shown a strong link between LMAF-related commensurate magnetic fluctuations at $\vec{Q}_0$ and
the PM--HO transition in \urusi. Temperature scans of the low-energy excitations at the
commensurate $\vec{Q}_0$ and the incommensurate $\vec{Q}_1$ positions show
qualitatively different behavior across the transition, with the former being strongly
enhanced towards the PM--HO transition temperature $T_0$. An analysis of energy scans in terms of damped harmonic oscillator functions characterized by the resonance energy $E_0$ and damping $D$ shows that the difference in the temperature scans originates from the temperature dependences of $1/\Gamma=D/E_0^2$ at $\vec{Q}_0$ and $\vec{Q}_1$. Our results put strong
constraints on theoretical models for the HO state; they point to a common, nearly isotropic,
order-parameter space involving both HO and LMAF
order parameters \cite{hau09b,cha13a}.

As a consequence we predict for high-pressure neutron scattering experiments of the
low-energy excitations near the PM--LMAF transition,
that the fluctuations at $\vec{Q}_0$ become stronger
for increasing pressure at $T_0(p)$, with a clear trend to a truly critical divergence at $T_N$
beyond the bicritical point \cite{mot03a,uem05a,has08a,mot08a,nik10a}.
At the same time, the intensity at $\vec{Q}_1$ is predicted to remain non-critical, step-like,
at all $p$ and $T$.

\section{Note added in proof}

We note two recent Raman spectroscopy reports (Refs. \onlinecite{kun15a} and \onlinecite{buh14a}) on the observation of a sharp resonance $A_{2g}$ mode at very similar energies as the $\vec{Q}_0$ excitations seen in neutron scattering. These works support our super-vector order parameter interpretation based on Refs. \onlinecite{hau09a} and \onlinecite{hau09b}.


\acknowledgments

We acknowledge fruitful discussions with P. B\"oni, P. Chandra, P. Coleman and R. Flint and also support by FRMII and by the German Science Foundation (DFG) through
FOR 960 (CP,MV), SFB/TR 80 (CP) and GRK 1621 (MV).

\vspace{-0.2cm}


\end{document}